# Smart Charging Impact Analysis using Clustering Methods and Real-world Distribution Feeders

Ravi Raj Shrestha
Department of Electrical and Computer Engineering
The University of Alabama
Tuscaloosa, Alabama, U.S.A
rrshrestha@crimson.ua.edu

Zhi Zhou, Limon Barua, Nazib Siddique, Karthikeyan Balasubramaniam, Yan Zhou
Energy Systems and Infrastructure Analysis
Argonne National Laboratory
Lemont, Illinois, U.S.A
{zzhou, lbarua, csiddique, kbalasubramaniam, yzhou}@anl.gov

Lusha Wang
Department of Electrical and Computer Engineering
The University of Alabama
Tuscaloosa, Alabama, U.S.A
lusha.wang@ua.edu

*Abstract*— The anticipated widespread adoption of electric vehicles (EVs) necessitates a critical evaluation of existing power distribution infrastructures, as EV integration imposes additional stress on distribution networks that can lead to component overloading and power quality degradation. Implementing smart charging mechanisms can mitigate these adverse effects and defer or even avoid upgrades. This study assesses the performance of two smart charging strategies—Time of Use (TOU) pricing and Load Balancing (LB)—on seven representative real-world feeders identified using k-means clustering. A time series-based steady-state load flow analysis was conducted on these feeders to simulate the impact of EV charging under both strategies across four different EV enrollment scenarios and three representative days to capture seasonal load characteristics. A grid upgrade strategy has been proposed to strengthen the power grid to support EV integration with minimal cost. Results demonstrate that both TOU and LB strategies effectively manage the additional EV load with reduced upgrade requirement and cost to existing infrastructure compared to the case without smart charging strategies and LB outperforms TOU when the customer enrollment levels are high. These findings support the viability of smart charging in facilitating EV integration while maintaining distribution network reliability and reducing investment cost.

## I. INTRODUCTION

Efforts to decarbonize transportation have surged EV adoption [1]. With the U.S. targeting 50% zero-emission vehicle sales by 2030 [2], the focus has shifted to building infrastructure for large-scale EV adoption. The IEA's 2030 forecast increased by 13–14% due to 2023 battery EV sales [3]. Assessing whether existing power distribution can handle such integration and proactively preparing grid by upgrading electrical components to avoid operational violations are essential.

The impact of EV charging on the power grid varies significantly across different feeders, as real-world feeders exhibit diverse characteristics in terms of feeder size, loading condition, capacity, and EV load profiles. EVs introduce additional loads to the existing base loads, and the impact largely depends on the timing and location of these charging loads. Charging during peak demand hours can place added stress on distribution lines and transformers, leading to reduced power quality and potential equipment degradation. Furthermore, charging load on small-sized transformers increases the risk of overloading. However, EV charging loads can also serve as flexible resources for the grid. By managing and coordinating charging schedules to align with base load profiles, it is often possible to mitigate adverse impacts and support grid reliability.

This study examines how utility-managed smart charging strategies can enhance overall system operations and benefit all participants in the smart charging program, particularly utilities and EV users. It involves (1) selecting representative feeders from utility distribution systems using k-means clustering, (2) converting representative feeder model from Cyme to OpenDSS for research purposes, (3) evaluating grid operational conditions of two smart charging strategies via time-series distribution flow analysis, (4) identifying upgrade requirement, time, and associated costs under various charging strategies and enrollment levels. The workflow is depicted on Fig. 1. The input data includes feeder configurations in OpenDSS format, hourly base load profiles generated using Advanced metering infrastructure (AMI) data, EV load profile at each feeder node, and estimated upgrade cost of each component in distribution systems. A Python environment is developed to execute load flow analyses and aggregate results. Distribution system details were provided by the utilities.

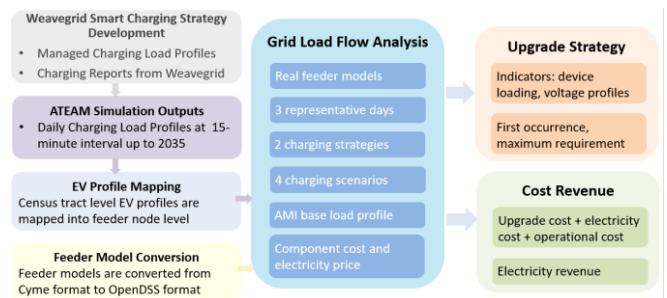

Fig. 1 Workflow diagram for the study

## II. PREVIOUS STUDIES ON IMPACT ASSESSMENT

### A. Feeder Clustering

Clustering is an unsupervised machine learning technique that groups similar data points while separating dissimilar ones [4]. It has been applied to identify representative feeders for PV integration analysis, with studies showing that feeders within the same cluster exhibit similar PV hosting capacities [5]–[7]. In [8], clustering was used to classify low-voltage feeders for network planning and hosting capacity studies. Similarly, [9] utilized K-means clustering to determine representative feeders for evaluating the impact of uncontrolled electric vehicle (EV) charging on distribution feeders. The authors used ratio of customer types and the associated aggregated loads as the features for clustering. Beyond hosting capacity determination, clustering is widely applied in areas such as fault detection [10] and reliability regulation [11], making it a versatile tool for grouping unlabeled data in various studies.

### B. Impact Analysis of EV Charging on Distribution Feeders

Several studies [9], [12]–[16] have assessed the impacts of integrating EVs into the distribution grids. Without smart charging strategies, feeder loading becomes a major concern, where uncontrolled charging can increase loads to as much as 15% above thermal conductor ratings [9]. Increased EV integration level can overload lines and transformers and reduce power quality, but these negative effects can be mitigated through controlled charging [12]–[15]. Reference [16] evaluated the impact of unmanaged EV charging integration on California feeders by comparing feeder capacity headroom with the total EV load at the substation level. The study concluded that 67% of the feeders will require capacity upgrades by 2045, with an estimated cost ranging from $6 to $20 billion. However, the feeder capacity headroom is an approximate estimate, and comparing these values directly overlooks the effects of EV charging location, timing, and base loads, resulting in a vague approximation that lacks accuracy. Detailed assessments of component overloading and voltage issues through load flow analysis within individual feeders are also not included.

Studies have shown that smart charging can significantly reduce the need for additional generation sources and infrastructure reinforcement [17-19]. In Great Britain, [20] found that with 100% EV adoption, smart charging reduced reinforcement requirements from 28% to 9%. This finding was based on simulations using three synthetic feeders representing urban, suburban and rural areas. The same feeder models were used for analysis of different geographic areas, with different traffic data applied to simulate regional variations. Similar studies [21], [22] reported that smart charging reduced transformer aging by 80% and loading by 20%, respectively. A financial analysis in [23] demonstrated that charging strategies minimizing both peak load and cost—especially those based on electricity prices—yield a better Net Present Value (NPV). The study emphasized that effective charging strategies must consider both cost reduction and peak load reduction. Additionally, a Shanghai-based study [24] concluded that time-of-use charging is a cost-efficient method for implementing smart charging. Nevertheless, most of the research discussed used standard or fictional distribution networks for analysis, leading to results that may not accurately reflect real-world operations. In our analysis, we address this gap by analyzing real-world representative distribution networks in the Maryland area (Fig. 2) to evaluate large-scale system-level impacts and estimate upgrade costs under various EV charging strategies. The analysis utilizes real distribution networks from Baltimore Gas and Electric (BGE) and Pepco Holdings Inc. (PHI), along with actual transportation data from the corresponding census tracts, enabling a more accurate assessment of system impact and upgrade requirements.

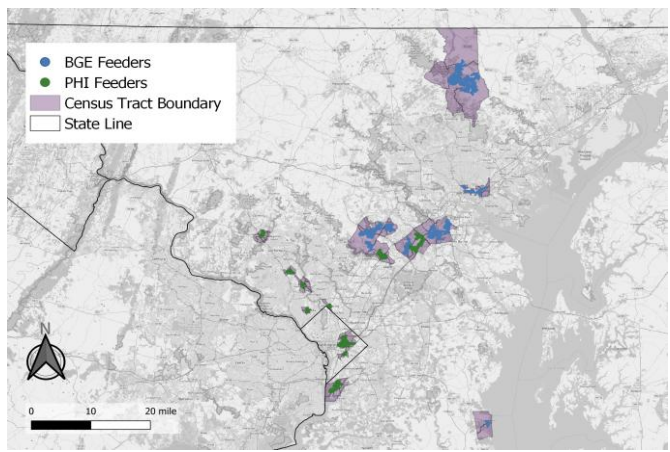

Fig. 2 Study geographic area

## III. APPLICATION OF K-MEANS FOR CLUSTERING

### A. Selection of Features and Number of Clusters

This study considered the key features that directly influence the generalization of the impact and cost results. These features include feeder voltage level, peak base load, peak EV load, total transformer capacity, counts of each transformer phase type, counts of each line phase type, and the total residential, commercial, industrial, and mixed-use loads. For clustering, we combined Principal Component Analysis (PCA) and K-Means: PCA transformed the input features into orthogonal components to reduce dimensionality and noise, while K-Means grouped the data into clusters based on similar features. To determine the optimal number of clusters, the Elbow Method was adopted, which indicated that seven clusters were most appropriate.

### B. Feeder Selection for Generalization

Clustering and analysis have been conducted for both BGE and PHI feeders, with this study primarily presenting the analysis results of BGE feeders. The clustering methodology groups 1,500 BGE feeders into seven clusters. Representative feeders were selected based on their proximity to the cluster centroid and analyzed to assess the effectiveness of two smart charging strategies in terms of grid upgrade needs. Fig. 3 presents the k-means clustering results of 1500 BGE feeders, including the decision boundaries, principal components and cluster centroid, while Table I provides a detailed description of each cluster's characteristics.

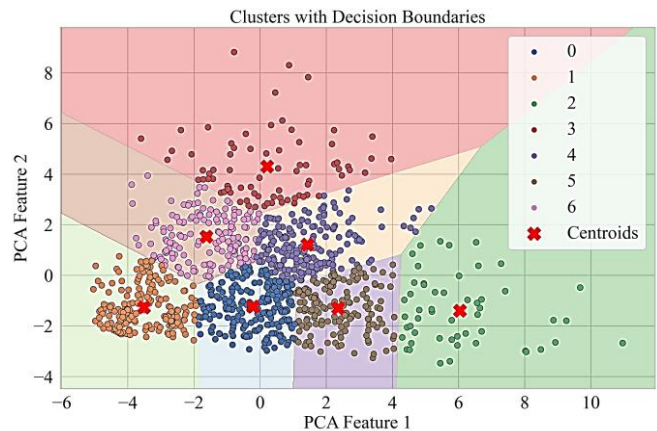

Fig. 3 Clustering results of 1500 BGE feeders

Table I Description of Each Cluster

| Cluster ID | Description |
| --- | --- |
| 0 | Moderate base and EV load, high transformer capacity, short feeder, mainly residential with some mixed-use. |
| 1 | Low base and EV load, high transformer capacity, very short feeder, residential with minimal commercial presence. |
| 2 | Very high base and EV load, very high capacity, longest feeder, serving significant residential, commercial, and mixed-use. |
| 3 | High base load, moderate EV load, high capacity, short feeder, primarily commercial with some residential and mixed-use. |
| 4 | High base and EV load, high capacity, moderate feeder, balanced residential, commercial, and mixed-use. |
| 5 | Moderate base and EV load, high capacity, moderate feeder, balanced residential and commercial loads. |
| 6 | Moderate base load, low EV load, moderate capacity, short feeder, primarily commercial with some industrial. |

## IV. EVALUATION OF SMART CHARGING STRATEGIES

After selecting representative feeders for each cluster, we evaluated two smart charging strategies—Time of Use (TOU) pricing and Load Balancing (LB)—to assess their effectiveness in terms of reducing the required upgrade investment. The TOU strategy reduces peak-hour charging demand by delaying EV charging between 5 p.m. and 9 p.m., resuming it afterward. Unlike unmanaged charging, where charging starts immediately upon plug-in, TOU temporarily pauses charging during peak hours, resuming after 9 p.m. This automation minimizes manual intervention, ensuring user convenience.

The LB strategy optimizes charging demand by grouping consumers based on grid assets and dynamically rescheduling sessions to prevent overloads. It gradually ramps up charging from evening to overnight, shifting loads to lower-demand periods. Using real-time monitoring and demand forecasting, it balances key parameters like load thresholds and user preferences to maintain grid stability while meeting charging needs. By combining these two approaches—shifting demand outside peak TOU windows and balancing the load across grid components—the overall system experiences reduced peak load, lower transformer overload risks, and potentially decreased upgrade costs. The EV charging load profile is generated using the Agent-Based Transportation Energy Analysis Model (ATEAM), which simulates EV user behaviors like plug-in/out times and charger locations[26, 27]. ATEAM assigns EVs to households using regional data and household travel surveys, modeling trips and charging sessions until the desired SOC or the next trip. It outputs detailed EV activity data, including charging times and power.

The study spanned from 2022 to 2035 and assessed each strategy under four scenarios: (1) no enrollment or uncontrolled charging, (2) minimum enrollment, where the rate reaches 10% by 2035; (3) steady increase in enrollment: where the rate reaches 30% by 2035; (4) maximum enrollment: where the rate gets to 50% by 2035. An upgrade strategy was formulated based on the loading status of electrical components as well as voltage profiles at each node. For example, if during the study period of 2022-2035, if the transformer is overloaded at year 2025, it will be upgraded in year 2025 with the capacity required to withstand the maximum loading of the entire study period. The upgrade plan also included installing voltage control components—such as capacitor banks and regulators—if voltage levels dropped below acceptable limits, after the overloading issues have been addressed. These components were added in the same year when under voltages were observed. The unit cost of components was based on [28]. Fig. 4 and Fig. 5 illustrate the effect of the two charging strategies on Feeder 1 load profiles across different enrollment levels. With a base peak load at 6 PM—indicating a residential area—both strategies effectively shifted EV charging o off-peak times. Under the LB strategy, the total peak load is reduced across all four scenarios. In contrast, under TOU strategy, although the original total peak load at 9 pm is reduced across all scenarios, the actual peak shifts to 5 pm in Scenario 3 and 4. This new peak occurs outside the TOU effective period, resulting in a second peak and offering no improvement in Scenario 3 and 4.

Table II compares performance of both smart charging strategies under Scenarios 1 and 4, with Fig. 6 illustrating transformer upgrade capacity needs. The LB strategy was generally more effective, achieving a 6% greater peak load reduction, 6% higher decrease in overloaded transformers, and 10% lower transformer upgrade costs than TOU. Small single-phase transformers were the most frequently overloaded, while most distribution lines had sufficient capacity, except for some low-capacity laterals.

The impact of different strategies varied by feeder. Feeder 6 showed no significant impact, as both strategies aligned with base load peaks. In Feeder 1, the strategies resulted in identical reductions in peak load and overloaded transformers; however, each strategy led to overloading in different transformers, necessitating varying upgrade capacities (kVA) and causing discrepancies in upgrade costs. For Feeder 7, under Scenario 4, the load shift caused several higher-rated transformers to become overloaded, increasing the total required upgrade capacity. Consequently, this led to higher transformer upgrade costs compared to Scenario 1, resulting in a negative cost reduction.

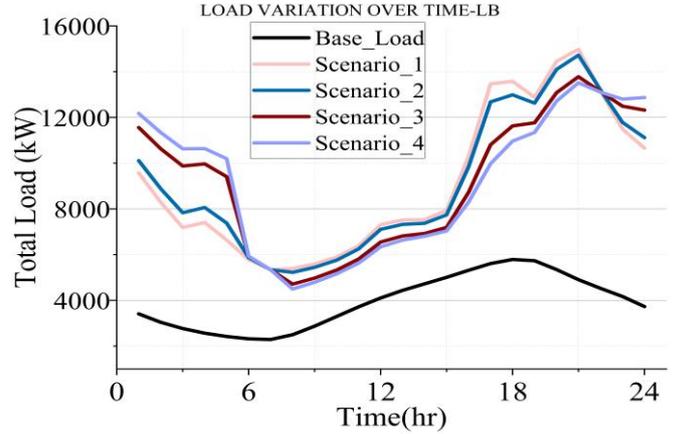

Fig. 4 Load profile of Feeder 1 with LB at year 2035

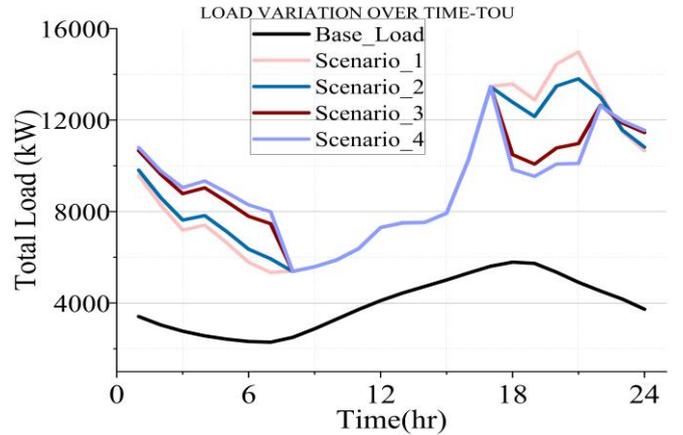

Fig. 5 Load profile of Feeder 1 with TOU at year 2035

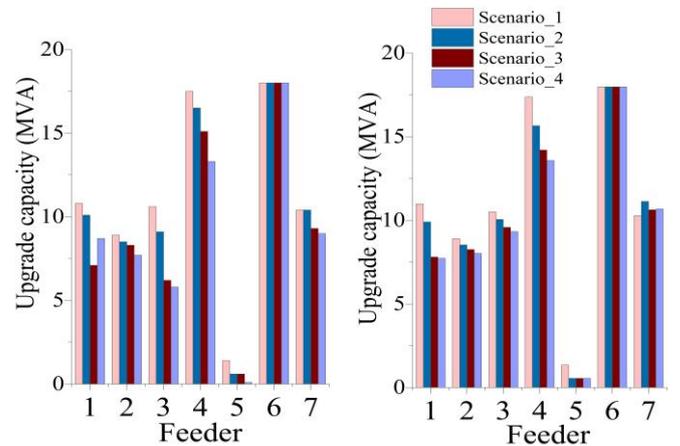

Fig. 6 Comparison of transformer upgrade capacity required for four scenarios of LB (left) and TOU (right)

Table II Comparison of Smart Charging Strategies(Scenario 1 vs Scenario 4)

| Cluster/Feeder | Strategy | Peak load reduction (%) | Overloaded transformer number reduction (%) | Transformer upgrade cost reduction (%) | Line upgrade cost reduction (%) |
|---|---|---|---|---|---|
| 0/1 | TOU | 10 | 23 | 34% | 3.5% |
|     | LB  | 10 | 23 | 20% | 28.5% |
| 1/2 | TOU | 8  | 8  | 8%  | 0% |
|     | LB  | 10 | 11 | 12% | 0% |
| 2/3 | TOU | 0  | 7  | 15% | 0% |
|     | LB  | 22 | 24 | 33% | 97.1% |
| 3/4 | TOU | 11 | 16 | 22% | 13.7% |
|     | LB  | 15 | 16 | 23% | 18.4% |
| 4/5 | TOU | 12 | 50 | 53% | 34.7% |
|     | LB  | 16 | 75 | 97% | 34.7% |
| 5/6 | TOU | 0  | 0  | 0%  | 0% |
|     | LB  | 2  | 0  | 0%  | 0% |
| 6/7 | TOU | 0  | 2  | -3% | 4.7% |
|     | LB  | 17 | 0  | 11% | 15.4% |

Fig.7 shows the cumulative overloaded transformer percentage across feeders from 2022 to 2035. While overloading increased under both strategies, higher smart charging enrollment reduced transformer overloading. Scenario 4, with maximum EV enrollment, had the fewest overloaded transformers, demonstrating smart charging's effectiveness. Though overloads remained unchanged from Scenario 3 to 4 under LB, delayed upgrades led to cost savings. Voltage violations occurred in three feeders, mainly on transformer secondary sides. Fig. 8 and Fig. 9 show voltage levels before and after upgrades. Transformer upgrades resolved undervoltage in one feeder, while the other two required capacitor banks.

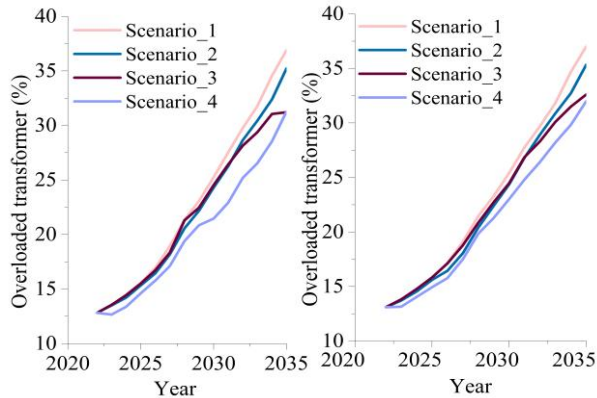

Fig. 7 Cumulative trend of overloaded transformers under LB (left) and TOU (right)

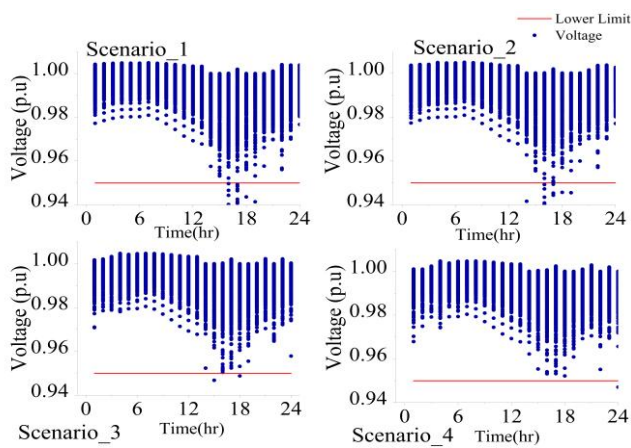

Fig. 8 Voltage profile for Feeder 7 before upgrade

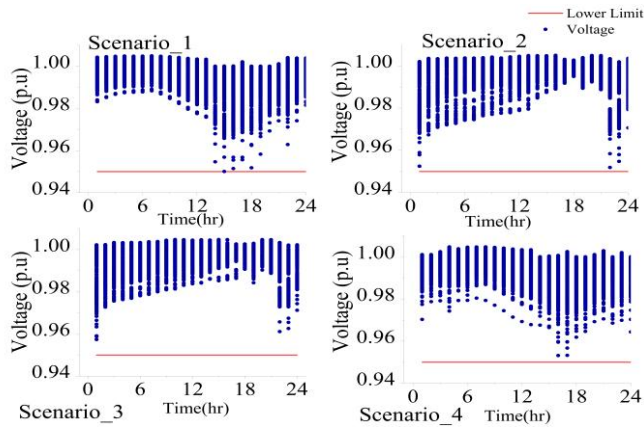

Fig. 9 Voltage profile for Feeder 7 after upgrade

The NPV of upgrade costs for LB and TOU strategies across the four scenarios is shown in Fig.10, calculated using a conservative 3% discount rate [25]. The analysis revealed that implementing smart charging strategies significantly lowers these costs across most feeders, with Scenario 4 being the most effective for future EV demands. This reduction in NPV suggests that smart charging can offset infrastructure upgrade expenses by optimizing energy usage and balancing peak and off-peak demand.

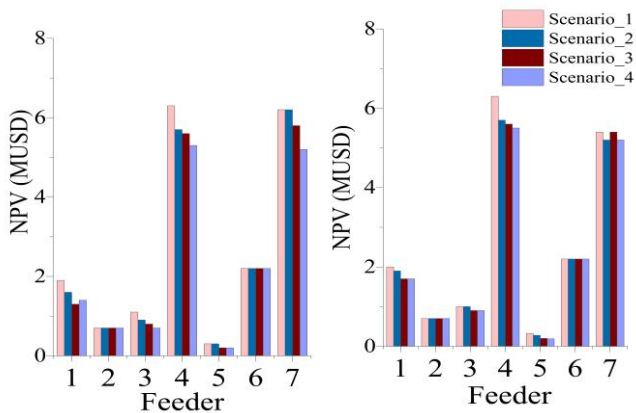

Fig 10. Upgrade cost for LB (Left) and TOU(Right) Scenarios

## V. Conclusion

With rising EV penetration, preparing distribution feeders for increased load is crucial. Unlike studies using synthetic models, this paper analyzes real Exelon feeders, projecting impacts until 2035. Clustering methods group similar feeders, reducing the need for extensive individual analysis..

A detailed study of 10 representative feeders shows that both TOU and LB smart charging mitigate uncoordinated EV charging effects, with impact and cost varying by feeder characteristics and enrollment levels. Transformers are most affected, with over 35% projected to be overloaded by 2035, while distribution line capacity remains sufficient and voltage issues minimal. Scenario 4, with the highest enrollment, provides the most efficient solutions and cost savings. LB-based strategy proves more effective, achieving 6% greater cost reduction than TOU.

## VI. Acknowledgment

This manuscript was prepared by UChicago Argonne, LLC, operator of Argonne National Laboratory, under DOE Contract No. DE-AC02-06CH11357, with support from the DOE Vehicles Technologies Office (VTO). The authors thank Stephanie Leach (BGE) and Joshua Cadoret (Pepco Holdings) for their guidance. The U.S. Government holds a nonexclusive,